\documentclass[twocolumn,amsmath,amssymb,prl,superscriptaddress]{revtex4-1}

\setcitestyle{super}

\usepackage[final,letterspace=-40]{microtype}
\usepackage[colorlinks=true, linkcolor=blue, urlcolor=blue,  citecolor=blue, anchorcolor=blue]{hyperref}

\usepackage{mathpazo} 

\usepackage{graphicx}
\usepackage{dcolumn}
\usepackage{bm}
\usepackage{helvet}

\usepackage{caption}
\DeclareCaptionLabelFormat{plain}{Figure #2 $\boldsymbol |$ }

\begin{document}

\title{Incoherent lensless imaging via coherency back-propagation}

\author{Ahmed El-Halawany}
\thanks{These two authors contributed equally}
\affiliation{CREOL, The College of Optics \& Photonics, University of Central Florida, Orlando, Florida 32816, USA}
\author{Andre Beckus}
\thanks{These two authors contributed equally}
\affiliation{Department of Electrical and Computer Engineering, University of Central Florida, Orlando, FL 32816, USA}
\author{H. Esat Kondakci}
\email{esat@creol.ucf.edu}
\author{Morgan Monroe}
\author{Nafiseh Mohammadian}
\affiliation{CREOL, The College of Optics \& Photonics, University of Central Florida, Orlando, Florida 32816, USA}
\author{George K. Atia}
\affiliation{Department of Electrical and Computer Engineering, University of Central Florida, Orlando, FL 32816, USA}
\author{Ayman F. Abouraddy}
\affiliation{CREOL, The College of Optics \& Photonics, University of Central Florida, Orlando, Florida 32816, USA}

\begin{abstract} \noindent 	
The two-point complex coherence function constitutes a complete representation for scalar quasi-monochromatic optical fields. Exploiting dynamically reconfigurable slits implemented with a digital micromirror device, we report on measurements of the complex two-point coherence function for partially coherent light scattering from a `scene' comprising one or two objects at different transverse and axial positions with respect to the source. Although the intensity shows no discernible shadows in absence of a lens,  numerically back-propagating the measured complex coherence function allows estimating the objects' sizes and locations -- and thus the reconstruction of the scene.
\end{abstract}

\small
\maketitle


\noindent 
The complex field amplitude $E(\mathbf{r})$ associated with a \textit{coherent} monochromatic scalar optical field provides a complete representation ($\mathbf{r}$ stands for the spatial coordinates) \cite{FundOfPhotonics}. Once the amplitude \textit{and} phase of $E(\mathbf{r})$ are measured -- by digital holography \cite{Yamaguchi97OL,Kim11Book}, acquiring the intensity in two planes \cite{Gerchberg72Optik,Fienup82AO,Abouraddy06NM,Witte14Light}, or wavefront sampling \cite{Vdovin15JO}, for example -- the field can be computed in any other plane using the diffraction propagator. When spatially \textit{incoherent} light scatters off an object, the far-field intensity no longer retains distinctive features. Although the transfer function representing free propagation of incoherent light has no zeros \cite{George97OC}, it nevertheless decays sharply with spatial frequency, thus significantly diminishing the contrast of far-field intensity variations and reducing the potential of identifying a scattering object. However, the two-point field correlations $G(\mathbf{r}_{1},\mathbf{r}_{2};\lambda)$ for pairs of points $\mathbf{r}_{1}$ and $\mathbf{r}_{2}$ in a quasi-monochromatic scalar field at a wavelength $\lambda$ provides a complete representation \cite{OpticalCoherenceBook}: determining $G(\mathbf{r}_{1},\mathbf{r}_{2};\lambda)$ at a plane allows evaluating it at any other plane. Measuring $G(\mathbf{r}_{1},\mathbf{r}_{2};\lambda)$ can be accomplished via wavefront sampling \cite{ThompsonPartiallyCoherent,Francon67PinO} or lateral-shear interferometry \cite{Iaconis96OL,Cheng00JMO}, among other possibilities \cite{Santarsiero06OL,Gonzales11JOSAA,DivittSunlight2015,Stoklasa14NC,Kagalwala15SR,Sharma16OE}. Other approaches to incoherent lensless imaging of an object include interferometric tomography \cite{Marks99Science} and rotational-shear interferometry \cite{Marks99AO}.

In this paper, we measure the coherence function of the optical field from an LED that is intercepted by a `scene' comprising one or more obstacles. The partially coherent field evolves after the scene until intensity variations representative of the objects (shadows) are no longer discernible. The coherence function is measured by implementing dynamically reconfigurable double slits \cite{CoherenceDMD} using a digital micromirror device (DMD) \cite{Dudley03SPIE}. For simplicity, we consider fields with one transverse coordinate $x$ (assuming all fields are uniform along the other coordinate) and obtain the magnitude  and phase of $G(x_{1},x_{2})$ at the detection plane from the visibility of the interferogram and the shift of the central fringe with respect to a fixed reference, respectively \cite{ThompsonPartiallyCoherent}, and then back-propagate $G(x_{1},x_{2})$ towards the source to discover the scene and locate the scattering objects (we drop $\lambda$ for convenience). We recently demonstrated that measuring $G$ along the $x_{2}\!=\!-x_{1}$ axis helps identify the transverse location and subtended angle (object width divided by its distance to the detection plane) of a single scattering object \cite{Kondakci17OE}. To identify the width and axial location separately, along with the transverse location, and -- furthermore -- to reconstruct a more complex scene, a measurement of the full coherence function becomes necessary -- as we proceed to demonstrate. 

\begin{figure*}[t!]\centering 
\includegraphics[scale=1]{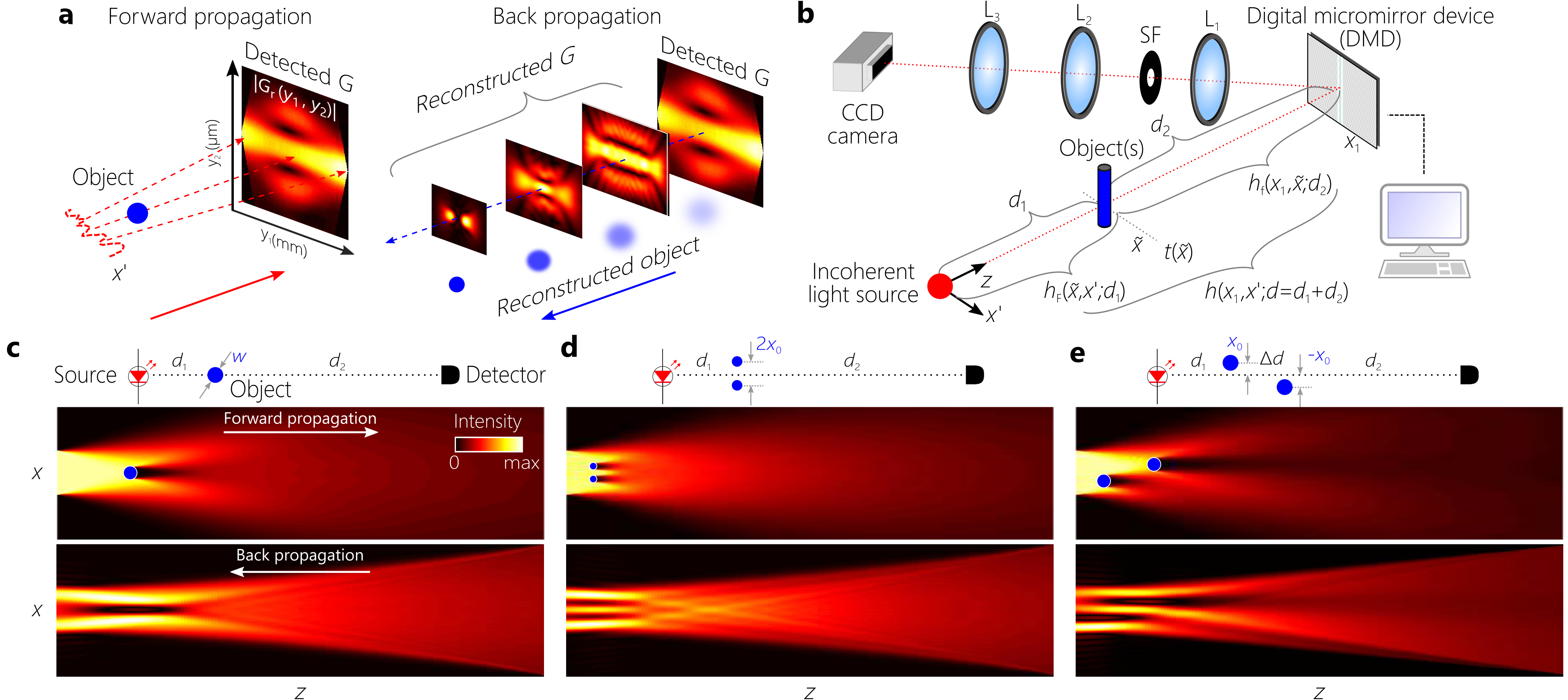}
 \caption{ (a) Concept of lensless coherence imaging. The coherence function $G(x_{1},x_{2};d)$ after scattering from an object is measured at a plane $z\!=\!d$, and then back-propagated computationally to the object. (b) Schematic of the measurement setup where relay lenses (L$_1$=10 cm and L$_2$=20~cm) are followed by a third lens in a $2f$ configuration (L$_3$=20~cm). SF: spatial filter. (c) A `scene' configuration comprising a single on-axis object with diameter $w\!=\!0.5$~mm, and $d_{1}\!=\!22$~cm. The distance between the source and detector plane $d\!=d_1\!+\!d_2\!=\!144$~cm is maintained throughout. In the forward direction, the object casts a shadow that washes out in the far-field. In the back-propagation direction, the object is replaced by an intensity dip that is symmetric with respect to the object location. (d) Configuration comprising two identical objects located in the same axial plane with $w\!=\!0.25$~mm, $x_{0}\!=\!0.287$~mm, and $d_{1}\!=\!7$~cm. (e) Configuration comprising two identical objects located in two different axial planes separated by a distance $\Delta d\!=\!15$~cm, with $w\!=\!0.5$~mm, $x_{0}\!=\!0.375$~mm, and $d_{1}\!=\!7$~cm. (c)-(e) In all simulations, $0\!\leq\!z\!\leq\!144$~cm and the $x$-axis spans 6~mm.}
\label{Fig:Concept}
\end{figure*}

The field correlations between two points $x_{1}$ and $x_{2}$ in a plane at a distance $z$ along the propagation axis can be described by a spatial coherence function $G(x_{1},x_{2};z)\!=\!\langle E(x_{1})E^{*}\!(x_{2})\rangle$, where $E(x)$ is a realization of a stochastic electric field and $\langle\cdot\rangle$ indicates an ensemble average, with the intensity laying along the diagonal $I(x_{1};z)\!=\!G(x_{1},x_{1};z)$. Starting from a planar source having a coherence function $G(x',x'';z\!=\!0)$, the coherence function at points $x_{1}$ and $x_{2}$ in a plane at $z\!=\!d$ after traversing a linear system having an impulse response function $h(x_{1},x')$ is
\begin{equation}\label{Eq:BasicPropagation}
G(x_{1},x_{2};d)\!=\!\!\!\int\!\!\!\!\int\!dx'dx''h(x_{1},x')h^{*}(x_{2},x'')G(x',x'';0). 
\end{equation}
Here, $h$ need not be unitary, so that systems including obstructions can be described in this way. In our experiments, $h$ comprises free-space propagation and interaction with opaque objects; see Fig.~\ref{Fig:Concept}. Propagation a distance $z$ is represented with a Fresnel integral of kernel $h_{\mathrm{F}}(x_{1},x';z)\!\propto\!\exp{\!\{i\tfrac{k}{2z}(x_{1}\!-\!x')^{2}\}}$ \cite{FundOfPhotonics}. In one configuration, $h$ comprises a sequence of free propagation a distance $d_{1}$ from the source, a thin opaque object represented by a transmittance $t(\tilde{x})$, followed by propagation a distance $d_{2}$ to the detection plane [Fig.~\ref{Fig:Concept}(a,b)]. This cascade is represented by the impulse response function
\begin{equation}
h(x_{1},x';d\!=\!d_{1}\!+\!d_{2})=\!\!\int\!d\tilde{x}\,\,h_{\mathrm{F}}(x_{1},\tilde{x};d_{2})\,\,t(\tilde{x})\,\,h_{\mathrm{F}}(\tilde{x},x';d_{1}),
\end{equation}
and the coherence function at the detector is
\begin{equation}\label{Eq:CoherenceWithObject}
G(x_{1}\!,x_{2};d)\!\!=\!\!\!\!\int\!\!\!\!\!\int\!\!d\tilde{x}d\tilde{\tilde{x}}h_{\mathrm{F}}(x_{1},\tilde{x};d_{2}\!)h_{\mathrm{F}}^{*}(x_{2},\tilde{\tilde{x}}\!;d_{2}\!)t(\tilde{x})t^{*}\!(\tilde{\tilde{x}}\!)G\!^{-}\!(\tilde{x},\tilde{\tilde{x}}\!;d_{1}\!), 
\end{equation}
where $G^{-}\!(\tilde{x},\tilde{\tilde{x}};d_{1})$ is the coherence function immediately before the object. We also define a coherence function immediately after the object $G^{+}\!(\tilde{x},\tilde{\tilde{x}};d_{1})\!=\!t(\tilde{x})t^{*}(\tilde{\tilde{x}})G^{-}\!(\tilde{x},\tilde{\tilde{x}};d_{1})$.

The specific form of the unitary operator for the Fresnel kernel $h_{\mathrm{F}}(x_{1},x;z)$ \cite{WhyIsLCDSoLittleKnown} leads to the identity $h_{\mathrm{F}}^{*}(x,x_{1};z)\!=\!h_{\mathrm{F}}(x,x_{1};-z)$ and a composition rule $\int\!\!d\tilde{x}h_{\mathrm{F}}(x_{1},\tilde{x};d_{2})h_{\mathrm{F}}(\tilde{x},x;d_{1})\!=\!h_{\mathrm{F}}(x_{1},x;d_{1}\!+\!d_{2})$. By setting $h_{\mathrm{F}}(x_{1},x;0)\!=\!\delta(x_{1}\!-\!x)$, $h_{\mathrm{F}}^{*}(x,x_{1};z)$ becomes the inverse of $h_{\mathrm{F}}(x_{1},x;z)$: $\int\!d\tilde{x}h_{\mathrm{F}}(x_{1},\tilde{x};z)h_{\mathrm{F}}^{*}(\tilde{x},x_{1};z)\!=\!\delta(x_{1}\!-\!x)$. Therefore, starting from the coherence function at the detector plane given in Eq.~\ref{Eq:CoherenceWithObject}, we can \textit{back-propagate} $G$ computationally a distance $z$ towards the object by applying the operator $h_{\mathrm{F}}^{*}(x,x_{1};z)\!=\!h_{\mathrm{F}}(x,x_{1};-z)$. When $z\!=\!d_{2}$, the back-propagated coherence function becomes $G^{+}\!(\tilde{x},\tilde{\tilde{x}};d_{1})$ and the intensity $I^{+}\!(\tilde{x};d_{1})\!=\!|t(\tilde{x})|^{2}I^{-}\!(\tilde{x};d_{1})$, where $I^{-}\!(\tilde{x};d_{1})$ is the intensity from the source immediately preceding the object.

Our strategy is thus to measure the complex coherence function and then carry out the back-propagation to reconstruct the scene. For convenience, we utilize a rotated coordinate system $(y_{1},y_{2})$ in lieu of $(x_{1},x_{2})$, where $y_{1}\!=\!(x_{1}\!+\!x_{2})/2$ and $y_{2}\!=\!(x_{1}\!-\!x_{2})/2$, such that $G(x_{1},x_{2})\!\rightarrow\!G_{\mathrm{r}}(y_{1},y_{2})$. This basis rotation helps visualize $G$ and facilitates computing its evolution along $z$ by enabling a higher sampling efficiency in the $(y_{1},y_{2})$-plane compared to that in the $(x_{1},x_{2})$-plane. The intensity lies along the $y_{1}$-axis $I(y_{1})\!=\!G_{\mathrm{r}}(y_{1},0)$, whereas the coherence properties are best gleaned along the $y_{2}$-axis.

The coherence function $G_{\mathrm{r}}(y_{1},y_{2})$ is measured via dynamically reconfigurable double-slits implemented by a DMD (Texas Instrument DLP 6500) that has $1024\!\times\!768$~pixels with a pixel pitch of 7.56~$\mu$m [Fig.~\ref{Fig:Concept}(b)]. The width of each slit is $\approx\!22.7$~$\mu$m (3~pixels). The separation between the slits $2y_{2}$ is varied in the range $0\!<\!2y_{2}\!<\!1500$~$\mu$m, whereas their center $y_{1}$ spans the range $-3\!<\!y_{1}\!<\!3$~mm with respect to the optical axis (DMD center) \cite{Kondakci17OE}. Following the DLP is a $4f$ imaging system (lenses L$_1$ and L$_2$ of focal lengths 10-cm and 20-cm, respectively, providing $2\!\times$ magnification) and a $2f$ system comprising a spherical lens L$_3$ of focal length 20~cm that produces interference patterns recorded by a CCD (The ImagingSource, DFK 31BU03), from which the magnitude and phase of $G_{\mathrm{r}}$ are extracted \cite{Kondakci17OE}; Fig.~\ref{Fig:Concept}(b). The source is incoherent light from an LED (Thorlabs M625L3, peak wavelength of $\approx\!633$~nm and FHWM-bandwidth $\approx\!18$~nm) filtered through a 1.3-nm-bandwidth filter at a wavelength of 632.8~nm.

The experiment is initially carried out in absence of objects (unobstructed propagation from the `primary' source to the detector) to calibrate the measurement system. The distance between the source and detector $d\!=\!1.44$~m is held fixed in all our experiments. We substitute the measured $G_{\mathrm{r}}(y_{1},y_{2};d)$ in the right-hand side of Eq.~\ref{Eq:BasicPropagation}, replace $h(x_{1},x)$ by $h_{\mathrm{F}}^{*}(x,x_{1},-z)$, and set the back-propagation distance to $z\!=\!d$. A calibration phase is assessed that produces a maximum intensity profile at the source plane of the back-propagated signal. We now proceed to reconstructing `secondary' sources -- the scattering objects.

\begin{figure}[t!]\centering 
\includegraphics[scale=1]{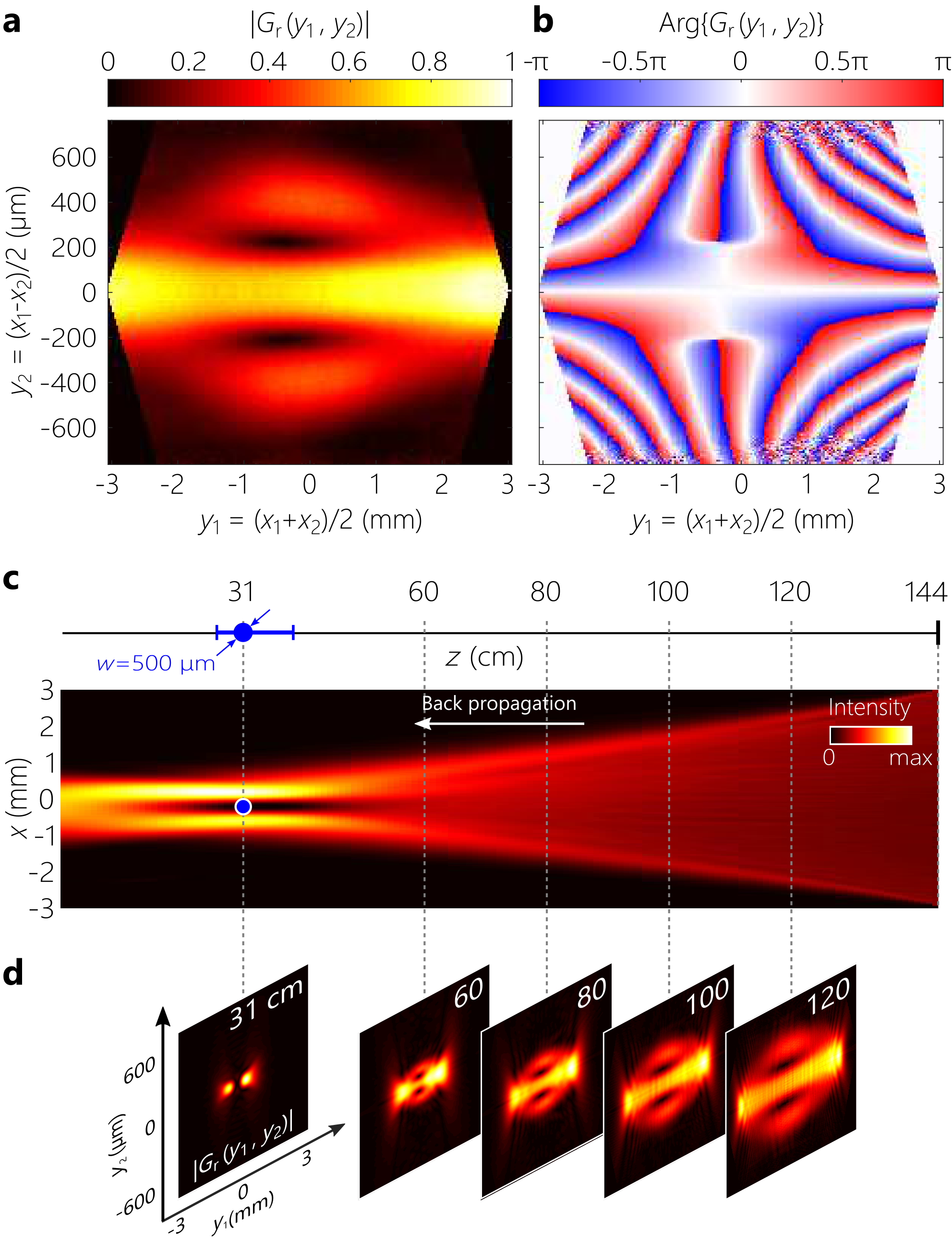}
  \caption{ Back-propagation of the measured coherence function for the configuration in Fig.~\ref{Fig:Concept}(c) comprising a single object. (a) Measured amplitude $|G_{\mathrm{r}}(y_{1},y_{2})|$ and (b) phase $\mathrm{Arg}\{G_{\mathrm{r}}(y_{1},y_{2})\}$ of the coherence function in the rotated coordinate system $(y_{1},y_{2})$. (c) Back-propagated intensity $I(x;z)$ along the $z$-axis. The object width is estimated to be 380~$\mu$m by finding the half-way point between the dip minimum and peak magnitude of the intensity in the object plane. The axial error bar indicates where the magnitude remains within $5\%$ of the minimum. (d) Back-propagated $G_{\mathrm{r}}(y_{1},y_{2};z)$ at selected distances from the source.}
\label{Fig:1Object}
\end{figure}

We first consider the case where an object (a metal wire of diameter $w\!=\!0.5$~mm at $d_{1}\!=\!22$~cm from the source) obstructs the field [Fig.~\ref{Fig:Concept}(c)]. Diffraction after the object smears out the shadow, as predicted by a forward-model calculation [Fig.~\ref{Fig:Concept}(c)] and confirmed in the measured $|G_{\mathrm{r}}(y_{1},0;d)|$ [Fig.~\ref{Fig:1Object}(a)]. By back-propagating the measured complex $G_{\mathrm{r}}(y_{1},y_{2};d)$ [Fig.~\ref{Fig:1Object}(a,b)] and increasing the back-propagation distance $z$, we construct the coherence function $G_{\mathrm{r}}(y_{1},y_{2};z)$ at planes preceding the detection plane axially and gradually approaching the object, samples of which are shown in Fig.~\ref{Fig:1Object}(d). From $G_{\mathrm{r}}(y_{1},y_{2};z)$ we can extract the evolution of the intensity distribution $I(y_{1};z)$ along the propagation axis by setting $y_{2}\!=\!0$ at every plane [Fig.~\ref{Fig:1Object}(d)]. Note the different scales along transverse (vertical) direction $x$ (4~mm) and longitudinal (horizontal) direction $z$ (1.44~m) in Fig.~\ref{Fig:1Object}(c).

The back-propagation yields a localized `shadow' of the object in the intensity profile that provides an estimate of the size and position (transverse and longitudinal) of the object [Fig.~\ref{Fig:1Object}(c)]. For simplicity, we consider the `focal plane' to be the plane in which the dip in the intensity profile reaches its minimum. The error in estimating the location of the object from the detection plane is $\approx\!7.4\%$. Note that the width of the intensity distribution decreases as we travel backwards and at the object plane is quite narrow; contrary to the extremely wide field produced from the LED. This is due to the finite size of the detection area: the source field far from the optical axis at the \textit{object} plane does not contribute to the \textit{detection} plane.

\begin{figure}[t!]\centering 
\includegraphics[scale=1]{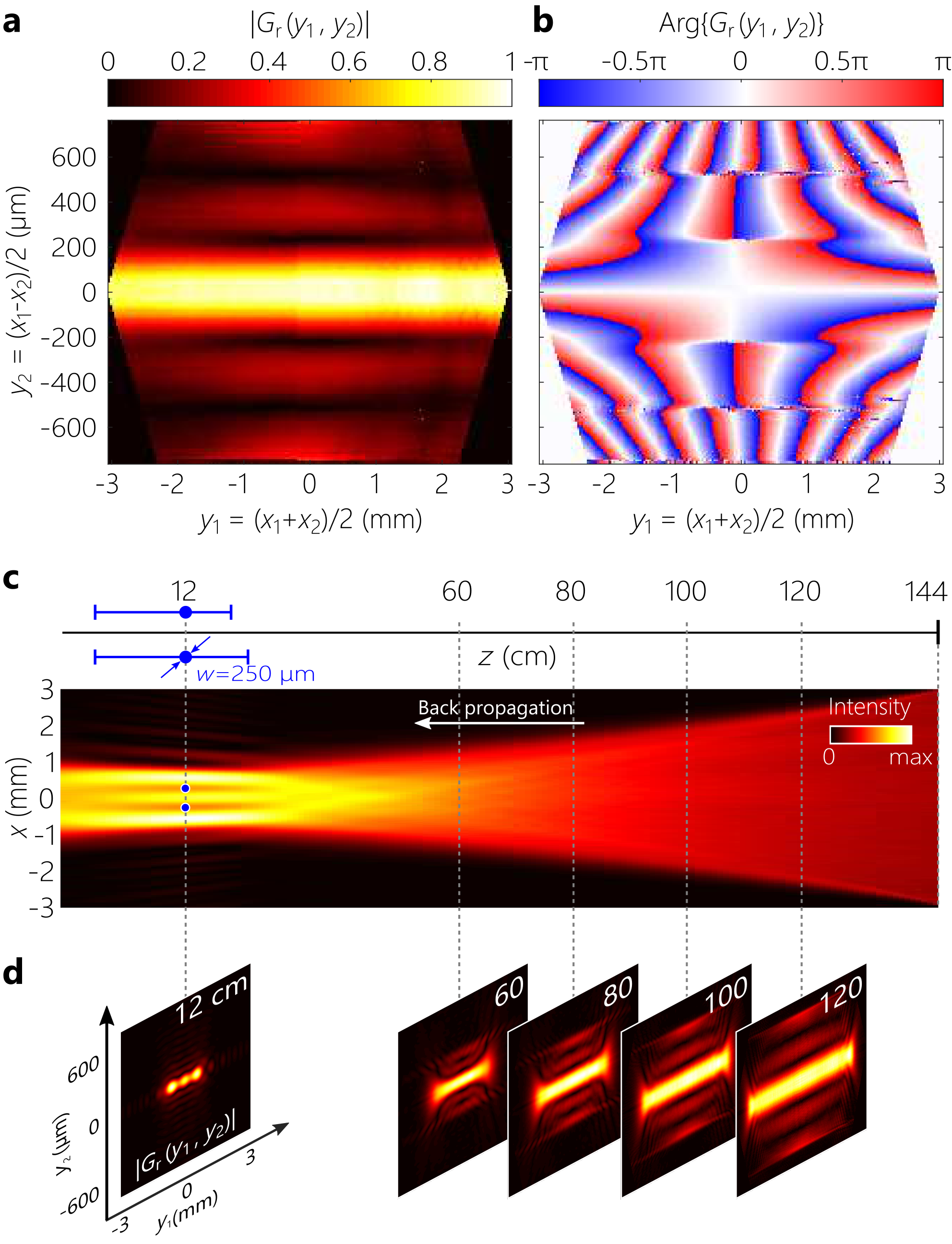}
\caption{Same as in Fig.~\ref{Fig:1Object}, except that the two-object scene in Fig.~\ref{Fig:Concept}(d) is employed. The two identical objects have a diameter of $250$~$\mu$m (smaller than that of the object in Fig.~\ref{Fig:1Object}) and are placed in the same transverse plane. The object widths are estimated at $\approx\!280$~$\mu$m at $d_{1}\!=\!12$~cm located at $x_{0}\!\approx\!-284$ and $\approx\!253$~$\mu$m.}
\label{Fig:2ObjectsSamePlane}
\end{figure}

We next consider a scenario where two co-planar objects: two metal wires of equal diameters $w\!=\!0.25$~mm separated by $2x_{0}\!=\!0.575$~mm and placed at a distance $d_{1}\!=\!7$~cm from the source [Fig.~\ref{Fig:2ObjectsSamePlane}(a)]. The shadow cast by the two objects has mostly smeared out at the detector plane; see $|G_{\mathrm{r}}(y_{1},0;d)|$ in Fig.~\ref{Fig:2ObjectsSamePlane}(a). The measured complex $G_{\mathrm{r}}(y_{1},y_{2};d)$ [Fig.~\ref{Fig:2ObjectsSamePlane}(a,b)] is back-propagated [Fig.~\ref{Fig:2ObjectsSamePlane}(d)], and we extract the evolution of the intensity $I(y_{1};z)$ along the propagation axis as before [Fig.~\ref{Fig:2ObjectsSamePlane}(c)]. The back-propagation yields two localized `shadows' of the objects in the intensity profile from which we estimate the size and position of the two objects [Fig.~\ref{Fig:2ObjectsSamePlane}(c)]. The error in estimating the location of the objects from the detection plane is $\approx\!3.6\%$.

\begin{figure}[t!]\centering 
\includegraphics[scale=1]{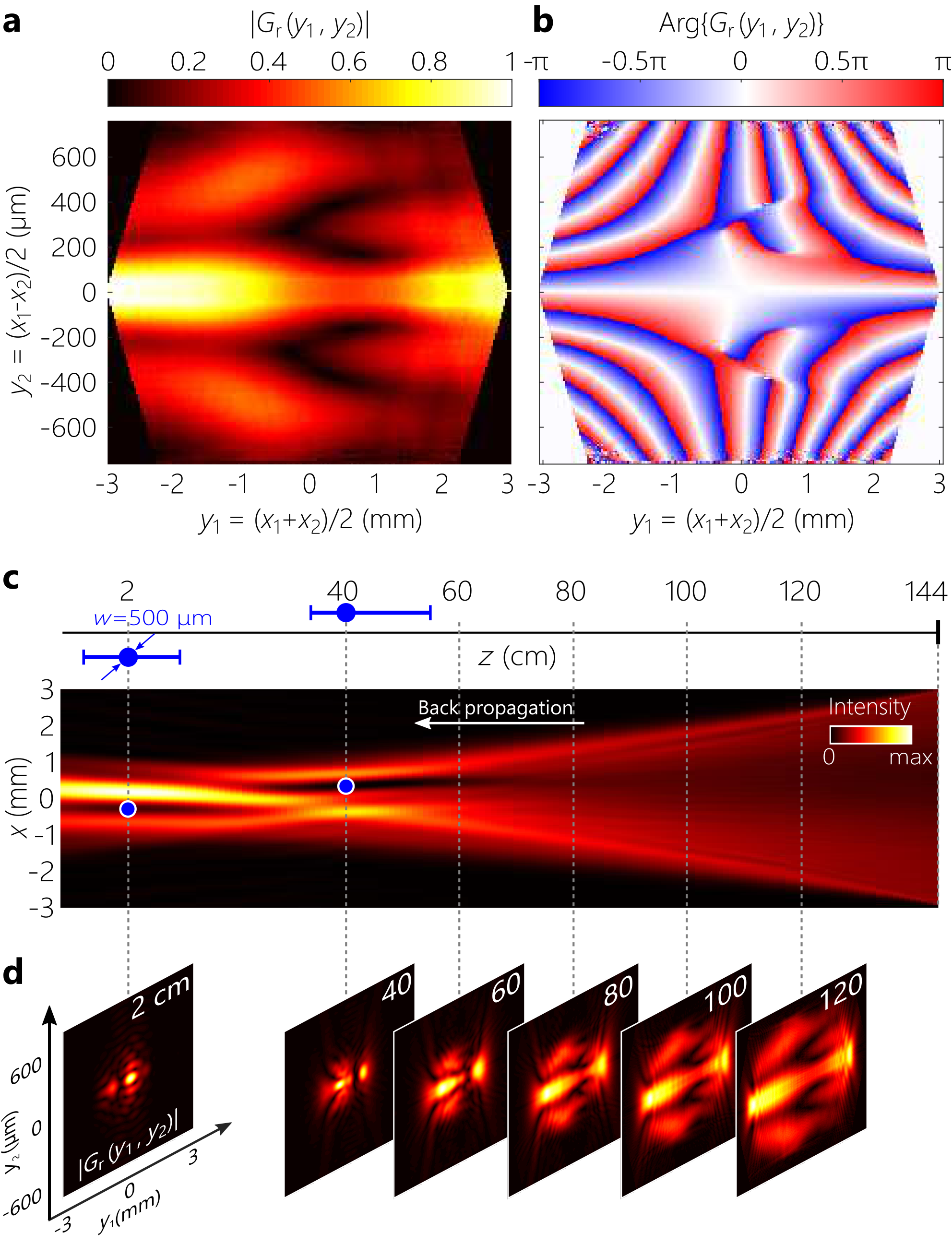}
\caption{Same as in Fig.~\ref{Fig:1Object}, except that the two-object scene in Fig.~\ref{Fig:Concept}(e) is employed. The two identical objects have a diameter of $500$~$\mu$m and are placed off the central axis in different axial planes. The object widths are estimated at $\approx\!400$~$\mu$m, $x_{0}\!\approx\!-321$~$\mu$m for the object closest to the source and $x_{0}\!\approx\!309$~$\mu$m for the object closest to the detector.}
\label{Fig:2ObjectsDiffPlaneSameSize}
\end{figure}

Finally, we consider a scenario where two objects (metal wires of diameter $w\!=\!0.5$~mm each) are located in \textit{different} planes along the propagation axis. The first object is at a distance $d_{1}\!=\!7$~cm from the source and is displaced to a position $x_{0}\!=\!0.375$~mm from the optical axis, and the second object follows it at a further distance $\Delta d$ along $z$ and is displaced to a symmetrically opposite transverse position $-x_{0}$ [Fig.~\ref{Fig:Concept}(e)]. Whereas the shadow cast by the first object (closest to the source) has mostly washed out, there is a remnant shadow from the second object (closest to the detection plane); see $|G_{\mathrm{r}}(y_{1},0;d)|$ in Fig.~\ref{Fig:2ObjectsDiffPlaneSameSize}(a). The measured complex $G_{\mathrm{r}}(y_{1},y_{2};d)$ [Fig.~\ref{Fig:2ObjectsDiffPlaneSameSize}(a,b)] is back-propagated [Fig.~\ref{Fig:2ObjectsDiffPlaneSameSize}(d)], and we extract the evolution of the intensity $I(y_{1};z)$ along the propagation axis as before [Fig.~\ref{Fig:2ObjectsDiffPlaneSameSize}(c)]. Over the course of the back-propagation, two localized `shadows' emerge. First, a shadow of the object closest to the detection plane emerges at $d_{1}\!=\!40$~cm in the intensity profile. We do not observe a shadow of the second object at this plane. By continuing the back-propagation procedure, the first observed shadow starts to smear out while a second shadow associated with the object closest to the source emerges. From these calculations, we can estimate the size and locations of the two objects [Fig.~\ref{Fig:2ObjectsDiffPlaneSameSize}(c)]. The errors in estimating the location of the objects from the detection plane are $\approx\!14.8\%$ and $\approx\!-3.6\%$.

We now discuss some of the limitations of this approach. The back-propagation is exact only if the detector is of infinite size. The finite detector size leads to imperfections in reconstructing the scene; e.g., a finite resolution for distinguishing objects located at neighboring transverse or longitudinal positions. The results in Fig.~\ref{Fig:1Object} identify a limitation of this approach, namely that the region immediately behind the object (which is occluded from the perspective of the detector) represents a `null space' for the procedure: a small object placed in the immediate vicinity behind the object will be difficult to observe. In general, when an object obstructs the light path, some information from the preceding planes is lost. For example, if two objects are placed in two different planes, the object closer to the detector will occlude the farther object. Finally, strictly speaking, the back-propagation procedure described above does not necessitate knowledge of the source for a successful reconstruction of the scene. We carried out a reference measurement for calibration only. An accurate measurement of $G_{\mathrm{r}}(y_{1},y_{2})$ suffices for the back-propagation procedure.

In conclusion, we have demonstrated that back-propagating the two-point complex coherence function measured in a plane can be utilized to reconstruct a scene containing scattering objects with no need for a lens. The coherence function need be measured at only one plane, even when the intensity in that plane lacks spatial variations indicative of the presence of objects. Acquiring the complex coherence function via wavefront sampling is made practical by utilizing a DMD that implements dynamically reconfigurable double slits. This work may be useful in imaging objects in turbid media where information lost in the intensity profile may still be glimpsed in the coherence domain.

\newpage 
\subsection*{Funding}
Defense Advanced Research Projects Agency (DARPA), Defense Science Office under contract HR0011-16-C-0029.
\vspace{-2mm}
\subsection*{Acknowledgments}
We thank D. Mardani for suggesting improvements to the  simulations. We also thank A. Tamasan, A. Dogariu, and A. Mahalanobis for helpful discussions.

%

\bibliography{main}
\end{document}